\documentstyle[12pt]{article}
\topmargin - 1.0 true cm
\newcommand{\beq}{\begin{equation}}
\newcommand{\eeq}{\end{equation}}
\def\sss{\scriptscriptstyle}

\def\bd{B_d^0}
\def\bdbar{{\overline{B_d^0}}}

\def\barp{{\raise.35ex\hbox{${\sss (}$}}---{\raise.35ex\hbox{${\sss )}$}}}
\def\bdbarp{\hbox{$B_d$\kern-1.4em\raise1.4ex\hbox{\barp}}}
\def\bsbarp{\hbox{$B_s$\kern-1.4em\raise1.4ex\hbox{\barp}}}
\def\ks{K_{\sss S}}
\def\dbar{{\overline{D^0}}}
\def\roughly#1{\mathrel{\raise.3ex\hbox{$#1$\kern-.75em\lower1ex\hbox{$\sim$}}}}

\def\bra#1{\langle#1|}
\def\ket#1{|#1\rangle}
\def\Dmbt{{\Delta m_{\sss B} \, t \over 2}}
\def\arnps#1#2#3{{\it Ann.\ Rev.\ Nucl.\ Part.\ Sci.} {\bf #1}, #3 (19#2)}
\def\epjc#1#2#3{{\it Eur.\ Phys.\ J.}\ {\bf C#1}, #3 (19#2)} 

\def\npb#1#2#3{{\it Nucl.\ Phys.} {\bf B#1}, #3 (19#2)}
\def\plb#1#2#3{{\it Phys.\ Lett.} {\bf #1B}, #3 (19#2)}
\def\prd#1#2#3{{\it Phys.\ Rev.} {\bf D#1}, #3 (19#2)}

\def\prl#1#2#3{{\it Phys.\ Rev.\ Lett.} {\bf #1}, #3 (19#2)}
\def\zpc#1#2#3{{\it Zeit.\ Phys.} {\bf C#1}, #3 (19#2)}

\begin{document}
\baselineskip=6truemm
\begin{flushright}
NSF-PT-99-5 \\
UdeM-GPP-TH-98-47 \\
\end{flushright}
\bigskip
\begin{center}
{\Large \bf 
Exploring CP Violation with $\bd\to D\ks$ Decays}
\bigskip\\
{\large Boris Kayser$^a$ and David London$^b$}
\end{center}

\bigskip

\begin{flushleft}
  ~~~~~~~~~~~$a$: {\it Division of Physics, National Science
    Foundation, 4201 Wilson Blvd.,}\\
  ~~~~~~~~~~~~~~~ {\it Arlington, VA 22230 USA}\\
  ~~~~~~~~~~~$b$: {\it Laboratoire Ren\'e J.-A. L\'evesque,
    Universit\'e de Montr\'eal,}\\
  ~~~~~~~~~~~~~~~{\it C.P. 6128, succ. centre-ville, Montr\'eal, QC,
    Canada H3C 3J7}
\end{flushleft}

\medskip 

\begin{center}
(\today)
\end{center}

\bigskip

\begin{quote}
{\bf Abstract}: 
We (re)examine CP violation in the decays $\bd\to D\ks$, where $D$
represents $D^0$, $\dbar$, or one of their excited states. The
quantity $\sin^2(2\beta + \gamma)$ can be extracted from the
time-dependent rates for $\bd(t)\to {\bar D}^{**0}\ks$ and $\bd(t)\to
D^{**0}\ks$, where the $D^{**0}$ decays to $D^{(*)+}\pi^-$. If one
considers a non-CP-eigenstate hadronic final state to which both
$\dbar$ and $D^0$ can decay (e.g.\ $K^+\pi^-$), then one can obtain
two of the angles of the unitarity triangle from measurements of the
time-dependent rates for $\bd(t) \to (K^+\pi^-)_{\sss D} \ks$ and
$\bd(t) \to (K^-\pi^+)_{\sss D} \ks$. There are no penguin
contributions to these decays, so all measurements are theoretically
clean.
\end{quote}
\newpage

\section{Introduction}

In the coming years, the CP-violating angles $\alpha$, $\beta$ and
$\gamma$ of the unitarity triangle will be measured in $B$ decays in a
number of different experiments \cite{CPreview}. The hope, as always,
is to find evidence of physics beyond the standard model (SM).

With this goal in mind, it is important to measure these three angles
in as many different ways as possible. There are (at least) two
reasons for this.  First, it is possible to discover new physics by
comparing values of the CP angles which are extracted in different $B$
decays. In fact, in this way one can often pinpoint this source of new
physics\footnote{For example, if the value of $\beta$ as extracted via
  the CP asymmetry in the decay $\bd\to\Psi\ks$ differs from that
  obtained in $\bd\to\phi\ks$, this indicates the presence of new
  CP-violating physics in the $b\to s$ penguin amplitude
  \cite{phiks}.}. Second, regardless of what decay mode is used, there
will always be some discrete ambiguities in the extraction of a CP
angle. These discrete ambiguities make it difficult to confirm (or
not) the predictions of the SM, and hence limit our ability to
discover new physics.  However, by using a variety of techniques, one
can measure different functions of the CP angles, which allows us to
remove the discrete ambiguities \cite{GrossQuinn,KayLon}.

In this paper, we (re)examine CP violation in the family of decays
$\bd\to D\ks$, where $D$ stands for $D^0$ or $\dbar$, as well as their
excited states. Since $\bd$ and $\bdbar$ mesons can each decay to both
$D^0$ and $\dbar$, this makes $D\ks$ final states a particularly rich
system to study.

For example, it has recently been pointed out that the weak phase
$2\beta+\gamma$ is probed in CP asymmetries in the decay $\bd\to
D^-\pi^+$ \cite{BDpi}. Here we show that this same phase can be
extracted from $\bd\to {\bar D}^{**0}\ks$, with the advantage that
roughly one third as many $B$'s are needed.

In fact, $\bd\to D\ks$ decays were studied many years ago
\cite{GroLon}.  Then it was shown that one could extract two of the
three angles of the unitarity triangle from the time-dependent rates
for $\bd\to D^0\ks$, $\bd\to\dbar\ks$ and $\bd\to D_{\sss CP}\ks$,
where $D_{\sss CP}$ denotes a $D^0$ or $\dbar$ decay to a CP
eigenstate. However, it was recently shown that this type of analysis
runs into problems because it is virtually impossible to tag the
flavor of the final-state $D$-meson \cite{ADS}, and so one cannot
distinguish $\bd\to D^0\ks$ from $\bd\to\dbar\ks$ decays. In this
paper we show that, despite these problems, it is still possible to
obtain two CP angles from a study of $\bd\to D^0\ks$ and
$\bd\to\dbar\ks$ if both $D^0$ and $\dbar$ decay to the same hadronic
final state (e.g.\ $K^+\pi^-$).

The paper is organized as follows. In Section 2 we examine how the CP
angle $2\beta + \gamma$ is extracted from both $\bd\to D^-\pi^+$ and
$\bd\to D\ks$ decays. Section 3 contains a discussion of how to obtain
two angles of the unitarity triangle (e.g.\ $\beta$ and $\gamma$) from
the time-dependent rates for $\bd(t) \to (K^+\pi^-)_{\sss D} \ks$ and
$\bd(t) \to (K^-\pi^+)_{\sss D} \ks$. We consider the question of
discrete ambiguities in Section 4. We conclude in Section 5.


\section{$\bd\to D\ks$: $2\beta + \gamma$}

It has been known for many years now that it is possible to cleanly
extract weak phase information using CP-violating rate asymmetries in
the $B$ system. The earliest studies of such rate asymmetries
concentrated on final states which are CP eigenstates. However, it
soon became clear that certain non-CP eigenstates can also be used. In
fact, as Aleksan, Dunietz, Kayser and Le Diberder (ADKL) showed
\cite{ADKL}, clean phase information can be obtained in $B$ decays to
almost any final state which is accessible to both $\bd$ and $\bdbar$.
We begin with a brief review of their method which, for the purposes
of identification, we will refer to later in the paper as the ADKL
method.

Consider a final state $f$ to which both $\bd$ and $\bdbar$ can decay.
Assume that $f$ is a two-body state, and that one weak amplitude
dominates both the $\bd$ and $\bdbar$ decays. (Both of these
conditions hold for the decays studied in this paper -- if one or both
of these conditions is not satisfied, then a more complicated analysis
is necessary.) We write
\begin{eqnarray}
\label{Bamps}
\bra{f} T \ket{\bd} = M e^{i\phi} e^{i\delta} ~~ & , & ~~~~
\bra{\bar f} T \ket{\bd} = {\overline M} e^{i{\bar\phi}} e^{i{\bar\delta}} ~,
\nonumber \\
\bra{\bar f} T \ket{\bdbar} = M e^{-i\phi} e^{i\delta} ~~ & , &
~~~~ \bra{f} T \ket{\bdbar} = {\overline M} e^{-i{\bar\phi}} e^{i{\bar\delta}}
~, 
\end{eqnarray}
where $\phi$ and ${\bar\phi}$ represent the weak phases of the decay,
and $\delta$ and ${\bar\delta}$ are the strong phases.

Due to $\bd$-$\bdbar$ mixing, a state which is created as a $\bd$ or a
$\bdbar$ will evolve in time into a mixture of both states:
\begin{eqnarray}
\ket{\bd(t)} & = & e^{-i m_{\sss B} t} e^{-\Gamma_{\sss B} t / 2} \left[
\cos\left(\Dmbt\right) \ket{\bd} - e^{- 2 i \phi_{\sss M}} i
\sin\left(\Dmbt\right) \ket{\bdbar} \right] ~, \nonumber \\
\ket{\bdbar(t)} & = & e^{-i m_{\sss B} t} e^{-\Gamma_{\sss B} t / 2} \left[
- e^{2 i \phi_{\sss M}} i \sin\left(\Dmbt\right) \ket{\bd} 
+ \cos\left(\Dmbt\right) \ket{\bdbar} \right] ~,
\label{timedep}
\end{eqnarray}
where $\phi_{\sss M}$ is the weak phase in $\bd$-$\bdbar$ mixing. (In
Eq.~(\ref{timedep}) the relative sign of the $\bd$ and $\bdbar$ terms
assumes, as indicated by lattice calculations, that the bag parameter,
$B_{\sss B_d}$, is positive. Even if this assumption is incorrect, the
analyses described below and in subsequent sections are largely,
though not totally, unaffected. We will make several comments
regarding the role of the bag parameter throughout the paper.) Using
the $B$-decay amplitudes defined in Eq.~(\ref{Bamps}), the
time-dependent decay rates for $\bd(t)$ and $\bdbar(t)$ to decay into
the final state $f$ become
\begin{eqnarray}
\label{Btofrates1}
\Gamma(\bd(t) \to f) & = & e^{-\Gamma_{\sss B} t} \left[ M^2
\cos^2\left(\Dmbt\right) + {\overline M}^2 \sin^2\left(\Dmbt\right) 
\right. \nonumber \\
& ~ & \qquad\qquad \left.
- M{\overline M} \sin(2 \phi_{\sss M} + \phi + {\bar\phi} + \delta -
{\bar\delta}) \sin(\Delta m_{\sss B} t) \right] ~, \\
\label{Btofrates2}
\Gamma(\bdbar(t) \to f) & = & e^{-\Gamma_{\sss B} t} \left[ {\overline M}^2
\cos^2\left(\Dmbt\right) + M^2 \sin^2\left(\Dmbt\right) 
\right. \nonumber \\
& ~ & \qquad\qquad \left.
+ M{\overline M} \sin(2 \phi_{\sss M} + \phi + {\bar\phi} + \delta -
{\bar\delta}) \sin(\Delta m_{\sss B} t) \right] ~, 
\end{eqnarray}
while those involving decays to ${\bar f}$ are
\begin{eqnarray}
\label{Btofbarrates1}
\Gamma(\bdbar(t) \to {\bar f}) & = & e^{-\Gamma_{\sss B} t} \left[ M^2
\cos^2\left(\Dmbt\right) + {\overline M}^2 \sin^2\left(\Dmbt\right) 
\right. \nonumber \\
& ~ & \qquad\qquad \left.
- M{\overline M} \sin(- 2 \phi_{\sss M} - \phi - {\bar\phi} + \delta -
{\bar\delta}) \sin(\Delta m_{\sss B} t) \right] ~, \\
\label{Btofbarrates2}
\Gamma(\bd(t) \to {\bar f}) & = & e^{-\Gamma_{\sss B} t} \left[ {\overline M}^2
\cos^2\left(\Dmbt\right) + M^2 \sin^2\left(\Dmbt\right) 
\right. \nonumber \\
& ~ & \qquad\qquad \left.
+ M{\overline M} \sin(- 2 \phi_{\sss M} - \phi - {\bar\phi} + \delta -
{\bar\delta}) \sin(\Delta m_{\sss B} t) \right] ~.
\end{eqnarray}

Through measurements of the above time-dependent rates\footnote{In
  fact, measurement of all four rates is not necessary. It suffices to
  measure one of the two rates in
  Eqs.~(\ref{Btofrates1})--(\ref{Btofrates2}), along with one of the
  two rates in Eqs.~(\ref{Btofbarrates1})--(\ref{Btofbarrates2}).}, it
is possible to extract the amplitudes $M$ and ${\overline M}$, as well
as the CP-violating quantities $S \equiv \sin(2\Phi + \Delta)$ and
${\bar S} \equiv \sin(2\Phi - \Delta)$, where $2\Phi \equiv 2
\phi_{\sss M} + \phi + {\bar\phi}$ and $\Delta \equiv \delta -
{\bar\delta}$. The two sines can be combined to yield
\beq
\sin^2 2\Phi = {1\over 2} \left[ 1 + S {\bar S} \pm \sqrt{(1-S^2)(1-{\bar
S}^2)} \right] ~.
\label{twosines}
\eeq
One of the signs gives the true value of $\sin^2 2\Phi$, while the
other gives $\cos^2 \Delta$. This discrete ambiguity can be removed by
repeating the analysis with another final state whose strong phases
are likely to be different. Thus, the ADKL method allows one to obtain
$\sin^2 2\Phi$ with no hadronic uncertainty.

Even if it turns out that, contrary to expectations, $B_{\sss B_d}$ is
in fact negative, the ADKL method will not be affected. The effect of
$B_{\sss B_d} < 0$ is to change the sign of the coefficient of the
$\sin(\Delta m_{\sss B} t)$ term in each of
Eqs.~(\ref{Btofrates1})--(\ref{Btofbarrates2}). In this case the
quantities $S$ and ${\bar S}$, as extracted from these rates, will
have the wrong sign. However, the weak phase $\sin^2 2\Phi$ obtained
from these quantities will be unaffected, since it depends only on the
products $S^2$, ${\bar S}^2$ and $S{\bar S}$ [Eq.~(\ref{twosines})].

It is amusing to note that that if the strong phase $\Delta$ is known
independently, $S$ and ${\bar S}$ can be combined to yield the weak
phase $2\Phi$ with no discrete ambiguity. However, if in fact $B_{\sss
  B_d} < 0$, then the weak phase obtained in this way will be $2\Phi +
\pi$. Thus, if one makes no assumption about the sign of $B_{\sss
  B_d}$, in this scenario there is a twofold ambiguity in the
extraction of $2\Phi$.

Recently, it has been noted that if one applies this technique to the
final state $f=D^-\pi^+$ (or $D^{*-}\pi^+$, $D^- \rho^+$, etc.), one
probes the weak phase $2\beta + \gamma$ \cite{BDpi}. This can be seen
as follows. The decays $\bd \to D^-\pi^+$ and $\bdbar \to D^-\pi^+$
are governed by the CKM matrix elements $V_{cb}^*V_{ud}$ and
$V_{ub}V_{cd}^*$, respectively. In the standard Wolfenstein phase
convention \cite{CPreview,Wolfenstein}, $\beta={\rm Arg}(V_{td}^*)$
and $\gamma={\rm Arg}(V_{ub}^*)$. Thus, for the final state $D^-\pi^+$
we have $\phi_{\sss M} = \beta$, $\phi = 0$, and ${\bar\phi} =
\gamma$. Therefore with the above method one extracts
$\sin^2\left(2\beta+\gamma\right)$.

One of the advantages of this method is that the branching ratios for
$\bd(t) \to f$ and $\bdbar(t) \to f$ ($f=D^-\pi^+$, $D^-\rho^+$, etc.)
are relatively large, in the range 3-8$\times 10^{-3}$. On the other
hand, the disadvantage is that the interfering amplitudes $\bd \to
D^-\pi^+$ and $\bdbar \to D^-\pi^+$ are quite different in size,
leading to a very small CP asymmetry:
\beq
A_{D\pi} \equiv {\Gamma(\bd(t) \to D^-\pi^+) - \Gamma(\bdbar(t) \to
D^+\pi^-) \over \Gamma(\bd(t) \to D^-\pi^+) + \Gamma(\bdbar(t) \to
D^+\pi^-)} \sim \left({{\overline M} \over M}\right)_{D\pi} \sim 
\left| {V_{ub}V_{cd}^* \over V_{cb}^*V_{ud}} \right| \sim 0.02 ~.
\eeq
This is a serious problem since the number of $B$'s needed to make the
measurement is inversely proportional to the square of the asymmetry:
\beq
\label{numberofBs}
N_{\sss B} \propto {1 \over BR(\bd\to f) \, A_f^2} ~.
\eeq
In Ref.~\cite{BDpi} it is estimated that one requires about $10^8$
tagged $B$'s to measure $|\sin(2\beta+\gamma)|$ to an accuracy of $\pm
0.1$.

One can in principle improve this situation by considering instead the
final state $f=\dbar\ks$. In this case the CKM matrix elements
involved in the decays $\bd \to \dbar\ks$ and $\bdbar \to \dbar\ks$
are $V_{cb}^*V_{us}$ and $V_{ub}V_{cs}^*$, respectively. (Technically,
we should also include the CKM matrix elements involved in
$K^0$-${\overline{K^0}}$ mixing. However, in the Wolfenstein
parametrization, these elements are real, and so do not contribute to
CP violation.) The phase information is unchanged compared to the
final state $D^-\pi^+$: $\phi_{\sss M} = \beta$, $\phi = 0$, and
${\bar\phi} = \gamma$. We therefore have the following time-dependent
decay rates:
\begin{eqnarray}
\label{BtoDbarKsrates1}
\Gamma(\bd(t) \to \dbar\ks) & = & e^{-\Gamma_{\sss B} t} \left[ M^2
\cos^2\left(\Dmbt\right) + {\overline M}^2 \sin^2\left(\Dmbt\right) 
\right. \nonumber \\
& ~ & \qquad\qquad \left. - M{\overline M} \sin(2 \beta + \gamma + \Delta) 
\sin(\Delta m_{\sss B} t) \right] ~, \\
\label{BtoDbarKsrates2}
\Gamma(\bdbar(t) \to \dbar\ks) & = & e^{-\Gamma_{\sss B} t} \left[ {\overline M}^2
\cos^2\left(\Dmbt\right) + M^2 \sin^2\left(\Dmbt\right) 
\right. \nonumber \\
& ~ & \qquad\qquad \left. + M{\overline M} \sin(2 \beta + \gamma + \Delta)
\sin(\Delta m_{\sss B} t) \right] ~, \\
\label{BtoDKsrates1}
\Gamma(\bdbar(t) \to D^0 \ks) & = & e^{-\Gamma_{\sss B} t} \left[ M^2
\cos^2\left(\Dmbt\right) + {\overline M}^2 \sin^2\left(\Dmbt\right) 
\right. \nonumber \\
& ~ & \qquad\qquad \left. - M{\overline M} \sin(- 2 \beta - \gamma + \Delta)
\sin(\Delta m_{\sss B} t) \right] ~, \\
\label{BtoDKsrates2}
\Gamma(\bd(t) \to D^0 \ks) & = & e^{-\Gamma_{\sss B} t} \left[ {\overline M}^2
\cos^2\left(\Dmbt\right) + M^2 \sin^2\left(\Dmbt\right) 
\right. \nonumber \\
& ~ & \qquad\qquad \left. + M{\overline M} \sin(- 2 \beta - \gamma + \Delta)
\sin(\Delta m_{\sss B} t) \right] ~.
\end{eqnarray}
Thus, with this final state, one can again extract
$\sin^2\left(2\beta+\gamma\right)$.

The advantage of using this final state is that, since the two
interfering amplitudes are of comparable size, the asymmetry is much
larger:
\beq
\label{MbarMsize}
A_{D\ks} \sim \left({{\overline M} \over M}\right)_{D\ks} \sim 
\left| {V_{ub}V_{cs}^* \over V_{cb}^*V_{us}} \right| \sim 0.4 ~.
\eeq
The disadvantage is that the branching ratios for $\bd(t) \to
\dbar\ks$ and $\bdbar(t) \to \dbar\ks$ are considerably smaller: we
estimate $B(\bd\to\dbar\ks) \approx \lambda^2 B(\bd\to\Psi\ks) = 2
\times 10^{-5}$. However, the net effect is still an improvement over
$\bd\to D^-\pi^+$. Although the branching ratio to $\dbar\ks$ is 150
times smaller than that to $D^-\pi^+$, the asymmetry is 20 times
bigger. From Eq.~(\ref{numberofBs}), we see that one therefore
requires roughly a factor of three fewer $B$'s to measure
$\sin^2(2\beta+\gamma)$ using the final state $\dbar\ks$.

Unfortunately, the final state $\dbar\ks$ (or $D^0\ks$) has its own
problems, namely that tagging the flavor of the final-state $D$-meson
is problematic \cite{ADS}. Attempts to tag the $\dbar$ via its
semileptonic decay $\dbar\to \ell^-{\bar\nu}_\ell X_{\bar s}$ are
hampered by the huge backgrounds from semileptonic $B$ decays. And
hadronic tags of the $\dbar$ such as $\dbar\to K^+\pi^-$ are not clean
either, since the $D^0$ can also decay to that final state, though the
amplitude is doubly Cabibbo-suppressed (DCS). We refer to this problem
as ``DCS contamination."

Suppose that one attempts to tag the flavor of the final-state
$D$-meson via the hadronic decay $\dbar\to K^+\pi^-$. What happens to
the expressions for the time-dependent rates for $\bd(t)$ and
$\bdbar(t)$ into $\dbar\ks$ and $D^0\ks$? We define
\begin{eqnarray}
\label{Damps}
\bra{K^-\pi^+} T \ket{D^0} = d \, e^{i\phi_d} e^{i\delta_d} ~~ & , & ~~~~
\bra{K^+\pi^-} T \ket{D^0} = {\bar d} \, e^{i{\bar\phi}_d}
e^{i{\bar\delta}_d} ~, \nonumber \\
\bra{K^+\pi^-} T \ket{\dbar} = d \, e^{-i\phi_d} e^{i\delta_d} ~~ & , &
~~~~ \bra{K^-\pi^+} T \ket{\dbar} = {\bar d} \, e^{-i{\bar\phi}_d}
e^{i{\bar\delta}_d} ~.
\end{eqnarray}
Note that, although these amplitudes have been written generally, in
fact we can set $\phi_d = {\bar\phi}_d = 0$ since the CKM matrix
elements involved in $D$ decays are essentially real in the
Wolfenstein parametrization. CLEO has measured \cite{CLEO}
\beq
\label{DKpirates}
{BR(D^0\to K^+\pi^-) \over BR(\dbar\to K^+\pi^-)} = 0.0077 \pm 0.0025 \pm
0.0025 ~.
\eeq
Taking the central value of this measurement, this gives
\beq
\label{dbardsize}
{{\bar d} \over d} \sim 0.09 ~. 
\eeq

Since $\bd$ and $\bdbar$ can decay into both $\dbar\ks$ and $D^0\ks$,
and since both $\dbar$ and $D^0$ can decay into $K^+\pi^-$, the decay
amplitudes must be added coherently, and the expressions for the
time-dependent $B$ decay rates become considerably more complicated:
\begin{eqnarray}
\label{BtoKpiKsrates1}
& ~ & \Gamma(\bd(t) \to (K^+\pi^-)_{\sss D} \ks) = 
e^{-\Gamma_{\sss B} t} \, e^{-\Gamma_{\sss D}t'} \times \nonumber \\
& ~ & \qquad\qquad\qquad \left\{
\cos^2\left(\Dmbt\right) \left[ M^2 d^2 + {\overline M}^2 {\bar d}^2 + 2 M
{\overline M} d {\bar d} \cos(\gamma - \Delta - \Delta_d) \right] \right.
\nonumber \\
&~& \qquad\qquad\qquad 
+ \sin^2\left(\Dmbt\right) \left[ {\overline M}^2 d^2 + M^2
{\bar d}^2 + 2 M {\overline M} d {\bar d} \cos(\gamma + \Delta -
\Delta_d) \right] \nonumber \\
&~& \qquad\qquad\qquad 
- \sin(\Delta m_{\sss B} t) \left[ 
M{\overline M} d^2 \sin(2 \beta + \gamma + \Delta)
+ M{\overline M} {\bar d}^2 \sin(2 \beta + \gamma - \Delta) \right. 
\nonumber \\
&~& \left. \left. \qquad\qquad\qquad\qquad\qquad
+ M^2 d {\bar d} \sin(2\beta + \Delta_d) 
+ {\overline M}^2 d {\bar d} \sin(2\beta + 2\gamma - \Delta_d) \right]
\right\} \\
\label{BtoKpiKsrates2}
& ~ & \Gamma(\bdbar(t) \to (K^+\pi^-)_{\sss D} \ks) =
e^{-\Gamma_{\sss B} t} \, e^{-\Gamma_{\sss D}t'} \times \nonumber \\
& ~ & \qquad\qquad\qquad \left\{
\cos^2\left(\Dmbt\right) \left[ {\overline M}^2 d^2 + M^2 {\bar d}^2 + 2 M
{\overline M} d {\bar d} \cos(\gamma + \Delta - \Delta_d) \right] \right.
\nonumber \\
&~& \qquad\qquad\qquad 
+ \sin^2\left(\Dmbt\right) \left[ M^2 d^2 + {\overline M}^2 
{\bar d}^2 + 2 M {\overline M} d {\bar d} \cos(\gamma - \Delta -
\Delta_d) \right] \nonumber \\
&~& \qquad\qquad\qquad 
+ \sin(\Delta m_{\sss B} t) \left[ 
M{\overline M} d^2 \sin(2 \beta + \gamma + \Delta) 
+ M{\overline M} {\bar d}^2 \sin(2 \beta + \gamma - \Delta)
\right. \nonumber \\
&~& \left. \left. \qquad\qquad\qquad\qquad\qquad
+ M^2 d {\bar d} \sin(2\beta + \Delta_d) 
+ {\overline M}^2 d {\bar d} \sin(2\beta + 2\gamma - \Delta_d) \right]
\right\} ,
\end{eqnarray}
where the $B$ and $D$ decays occur at times $t$ and $t'$, respectively
($\Gamma_{\sss D}$ is the $D$ width), with $\Delta \equiv \delta -
{\bar\delta}$ and $\Delta_d \equiv \delta_d - {\bar\delta}_d$.  The
rates for the CP-conjugate processes are obtained simply by changing
the signs of the weak phases:
\begin{eqnarray}
\label{BtoKpiKsbarrates1}
& ~ & \Gamma(\bdbar(t) \to (K^-\pi^+)_{\sss D} \ks) =
e^{-\Gamma_{\sss B} t} \, e^{-\Gamma_{\sss D}t'} \times \nonumber \\
& ~ & \qquad\qquad\qquad \left\{
\cos^2\left(\Dmbt\right) \left[ M^2 d^2 + {\overline M}^2 {\bar d}^2 + 2 M
{\overline M} d {\bar d} \cos(-\gamma - \Delta - \Delta_d) \right] \right.
\nonumber \\
&~& \qquad\qquad\qquad 
+ \sin^2\left(\Dmbt\right) \left[ {\overline M}^2 d^2 + M^2
{\bar d}^2 + 2 M {\overline M} d {\bar d} \cos(-\gamma + \Delta -
\Delta_d) \right] \nonumber \\
&~& \qquad\qquad\qquad 
- \sin(\Delta m_{\sss B} t) \left[ 
M{\overline M} d^2 \sin(- 2 \beta - \gamma + \Delta)
+ M{\overline M} {\bar d}^2 \sin(- 2 \beta - \gamma - \Delta) \right. 
\nonumber \\
&~& \left. \left. \qquad\qquad\qquad\qquad\qquad
+ M^2 d {\bar d} \sin(- 2\beta + \Delta_d) 
+ {\overline M}^2 d {\bar d} \sin(- 2\beta - 2\gamma - \Delta_d) \right]
\right\} \\
\label{BtoKpiKsbarrates2}
& ~ & \Gamma(\bd(t) \to (K^-\pi^+)_{\sss D} \ks) =
e^{-\Gamma_{\sss B} t} \, e^{-\Gamma_{\sss D}t'} \times \nonumber \\
& ~ & \qquad\qquad\qquad \left\{
\cos^2\left(\Dmbt\right) \left[ {\overline M}^2 d^2 + M^2 {\bar d}^2 + 2 M
{\overline M} d {\bar d} \cos(-\gamma + \Delta - \Delta_d) \right] \right.
\nonumber \\
&~& \qquad\qquad\qquad 
+ \sin^2\left(\Dmbt\right) \left[ M^2 d^2 + {\overline M}^2 
{\bar d}^2 + 2 M {\overline M} d {\bar d} \cos(-\gamma - \Delta -
\Delta_d) \right] \nonumber \\
&~& \qquad\qquad\qquad 
+ \sin(\Delta m_{\sss B} t) \left[ 
M{\overline M} d^2 \sin(- 2 \beta - \gamma + \Delta) 
+ M{\overline M} {\bar d}^2 \sin(- 2 \beta - \gamma - \Delta)
\right. \nonumber \\
&~& \left. \left. \qquad\qquad\qquad\qquad\qquad
+ M^2 d {\bar d} \sin(- 2\beta + \Delta_d) 
+ {\overline M}^2 d {\bar d} \sin(- 2\beta - 2\gamma - \Delta_d) \right]
\right\} .
\end{eqnarray}
Note that, in the limit where ${\bar d} \to 0$ and $d \to 1$ (i.e.\ no
doubly Cabibbo suppressed $D$ decays), the above equations reduce to
those of Eqs.~(\ref{BtoDbarKsrates1})--(\ref{BtoDKsrates2}).

{}From these expressions, it is clear that the fact that one cannot
cleanly tag the final $D$ meson introduces a significant uncertainty
into the extraction of $\sin^2(2\beta+\gamma)$. For example, if there
were no DCS contamination, the quantities $M{\overline M} \sin(2\beta
+ \gamma + \Delta)$ and $M{\overline M} \sin(2\beta + \gamma -
\Delta)$ could be respectively extracted from the coefficients of the
$\sin(\Delta m_{\sss B} t)$ terms in Eqs.~(\ref{BtoDbarKsrates1}) and
(\ref{BtoDKsrates2}). However, from Eqs.~(\ref{BtoKpiKsrates1}) and
(\ref{BtoKpiKsbarrates1}) above, we see that the presence of DCS
contamination introduces an uncertainty in the extraction of these
quantities:
\beq
\label{2betagammaerrors}
{\Delta\left( M{\overline M} \sin(2\beta + \gamma + \Delta) \right)
\over 
M{\overline M} \sin(2\beta + \gamma + \Delta)}
 \sim
{\Delta\left( M{\overline M} \sin(2\beta + \gamma - \Delta)\right) 
\over
M{\overline M} \sin(2\beta + \gamma - \Delta)}
\sim
{M^2 d {\bar d} \over M {\overline M} d^2} \sim 22\% ~,
\eeq
where we have used the estimates for ${\overline M}/M$ and ${\bar
  d}/d$ given in Eqs.~(\ref{MbarMsize}) and (\ref{dbardsize}). (Here
and in the following equation, the symbol $\Delta$ used to indicate
the error should not be confused with the same symbol which denotes
the strong phase $\delta - {\bar\delta}$.) Furthermore, via a similar
analysis, the extraction of amplitudes $M$ and ${\overline M}$ also
has errors induced:
\begin{eqnarray}
\label{MMbarerrors}
{\Delta M \over M} & \sim & 
{M {\overline M} d {\bar d} \over M^2 d^2} \sim 4\%
\nonumber \\
{\Delta {\overline M} \over {\overline M} } & \sim & 
{M {\overline M} d {\bar d} \over {\overline M}^2 d^2} \sim 22\% ~.
\end{eqnarray}
Clearly, when all these errors are put together, there is a
considerable systematic error in the extraction of the quantity
$\sin^2(2\beta + \gamma)$. Thus, the above analysis shows that, in
fact, due to the problems of tagging the final $\dbar$ meson, the
final state $\dbar\ks$ cannot be used to cleanly obtain
$\sin^2(2\beta+\gamma)$ via the ADKL method.  (Nevertheless, we can
learn a great deal from the decays of $\bd(t)$ and $\bdbar(t)$ to
$D^0\ks$ and $\dbar\ks$ in a different way, as we will show in the
next section.)

The problems with DCS contamination can be avoided if one uses instead
a final state involving a self-tagging excited $D^0$ state such as
$D_1(2420)^0$ or $D_2^*(2460)^0$ \cite{PDG}, generically denoted as
$D^{**0}$. The $D^{**0}$ decays to $D^{(*)+}\pi^-$, while the
CP-conjugate state decays to $D^{(*)-}\pi^+$. The charge of the pion
therefore tags the flavor of the decaying $D$-meson. Thus, the ADKL
method can be used with the final state $D^{**0}\ks$ or ${\bar
  D}^{**0}\ks$ to extract $\sin^2(2\beta + \gamma)$. This final state
has no problems with DCS contamination, so the measurement is clean.

It is also possible in principle to use three-body final states such
as $D^+\pi^-\ks$, $D_s^+K^-\ks$, etc.\ in order to obtain
$\sin^2(2\beta + \gamma)$. However, there is a problem: such states
will have nontrivial kinematic degrees of freedom due to the fact that
the relative angular momenta of the final-state particles are not
fixed. The ADKL method then applies only to a specific kinematical
point, or in a small kinematical bin. Since one requires a huge number
of $B$'s in order to accumulate an appreciable number of events in a
small bin, the application of the ADKL method to such three-body final
states is likely to be impractical \cite{PASCOS}.

Finally, we note that a measurement of $\sin^2(2\beta + \gamma) =
\sin^2(\beta-\alpha)$ does not, by itself, give any information about
the angles $\alpha$, $\beta$ and $\gamma$. However, if $\beta$ is
measured in another $B$ decay (e.g.\ in $\bd(t), \bdbar(t) \to \Psi
\ks$), then this information can be used in order to obtain $\alpha$
or $\gamma$, up to discrete ambiguities. And if two of the CP angles
are measured elsewhere, then this method serves as an independent
crosscheck. We will have more to say about this in section 4.


\section{$\bd\to D\ks$: $\beta$, $\gamma$}

Gronau and London (GL) suggested another method for obtaining clean
weak phase information from non-CP-eigenstate final states
\cite{GroLon}. It involves the decays $\bd\to D^0\ks$,
$\bd\to\dbar\ks$ and $\bd\to D_{\sss CP}\ks$, where $D_{\sss CP}$ is
the CP-even superposition
\beq
D_{\sss CP} = {1\over \sqrt{2}} \left[ D^0 + \dbar \right].
\eeq
$D_{\sss CP}$ is identified by its decays to CP-even final states such
as $\pi^+\pi^-$, $K^+K^-$, etc. (One can also use the CP-odd
combination of $D^0$ and $\dbar$ -- the CP asymmetry simply has an
extra minus sign.)

The GL method works as follows. From the time-dependent rates for
$\bd(t)$ to decay into $\dbar\ks$ and $D^0\ks$
[Eqs.~(\ref{BtoDbarKsrates1}) and (\ref{BtoDKsrates2})], one can
extract the quantities $M$, ${\overline M}$, $\sin(2\beta + \gamma -
\Delta)$ and $\sin(2\beta + \gamma + \Delta)$, as before. But there is
important, additional information to be obtained by considering also
the decay $\bd(t)\to D_{\sss CP}\ks$. The time-dependent rate is given
by
\begin{eqnarray}
& ~ & \Gamma(\bd(t) \to D_{\sss CP} \ks) = 
{1\over 2} e^{-\Gamma_{\sss B} t} \, e^{-\Gamma_{\sss D}t'} \times 
\nonumber \\
& ~ & \qquad\qquad\qquad \left\{
\cos^2\left(\Dmbt\right) \left[ M^2 + {\overline M}^2 + 2 M
{\overline M} \cos(\gamma - \Delta) \right] \right.
\nonumber \\
&~& \qquad\qquad\qquad 
+ \sin^2\left(\Dmbt\right) \left[ {\overline M}^2 + M^2
+ 2 M {\overline M} \cos(\gamma + \Delta) \right] \nonumber \\
&~& \qquad\qquad\qquad 
- \sin(\Delta m_{\sss B} t) \left[ 
M^2 \sin(2\beta) + {\overline M}^2 \sin(2\beta + 2\gamma) 
\right. 
\nonumber \\
&~& \left. \left. \qquad\qquad\qquad\qquad\qquad
+ M{\overline M} \sin(2 \beta + \gamma + \Delta)
+ M{\overline M} \sin(2 \beta + \gamma - \Delta) 
\right]
\right\} ~.
\end{eqnarray}
The measurement of this rate yields the additional quantities
$\cos(\gamma-\Delta)$ and $\cos(\gamma+\Delta)$.

{}From these four trigonometric quantities --- $\sin(2\beta + \gamma -
\Delta)$, $\sin(2\beta + \gamma + \Delta)$, $\cos(\gamma-\Delta)$ and
$\cos(\gamma+\Delta)$ --- it is straightforward to show that one can
obtain $\sin(2\beta)$ and $\sin(2\beta + 2\gamma) = -\sin(2\alpha)$.
Thus, two angles of the unitarity triangle can in principle be
extracted with no hadronic uncertainty from the time-dependent
measurements of $\bd(t) \to \dbar\ks$, $D^0\ks$, and $D_{\sss CP}\ks$.

This technique was adapted by Gronau and Wyler (GW) to the decays
$B^\pm \to \dbar K^\pm$, $D^0 K^\pm$ and $D_{\sss CP} K^\pm$ as a
probe of the angle $\gamma$ \cite{GroWyler}. However, it was recently
pointed out by Atwood, Dunietz and Soni (ADS) that this method runs
into the problems of DCS contamination mentioned in the previous
section \cite{ADS}.  Specifically, although the branching ratio for
$B^+\to \dbar K^+$ can be measured, obtaining $B(B^+\to D^0 K^+)$ is
extremely difficult due to the problems of tagging the final state
$D$-meson. Nevertheless, ADS were able to save the GW method. They
pointed out that one can still obtain the angle $\gamma$, up to a
fourfold discrete ambiguity, by studying decays such as $B^+ \to (K^+
\pi^-)_{\sss D} K^+$ and $B^+ \to (K^+ \rho^-)_{\sss D} K^+$, along
with their CP-conjugates. Note that final states involving $D_{\sss
  CP}$ are not necessary.

This raises the following questions. First, is the GL method affected
by DCS contamination? And second, if so, can it be rescued in a
similar manner to the ADS modification of the GW method?

The answer to both of these questions is yes. Including the DCS
contamination, the time-dependent decay rates of $\bd(t)$ into
$\dbar\ks$ and $D^0\ks$ are given in the previous section in
Eqs.~(\ref{BtoKpiKsrates1})--(\ref{BtoKpiKsbarrates2}). As discussed
there, due to DCS contamination, it is {\it not} possible to obtain
the quantities $\sin(2\beta + \gamma - \Delta)$ and $\sin(2\beta +
\gamma + \Delta)$ precisely [see Eq.~(\ref{2betagammaerrors})], so the
GL method breaks down.

Fortunately, the method can be saved in a fashion analogous to the ADS
modification of the GW method. Referring again to
Eqs.~(\ref{BtoKpiKsrates1})--(\ref{BtoKpiKsbarrates2}), we make the
following two observations. First, these four time-dependent rates
depend on four amplitudes ($M$, ${\overline M}$, $d$, ${\bar d}$) and
four phases ($\gamma$, $\beta$, $\Delta$, $\Delta_d$). Of these eight
quantities, the two amplitudes $d$ and ${\bar d}$ have been measured
[Eq.~(\ref{DKpirates})].  Second, because of the time dependence, six
independent quantities can be extracted from the measurements of these
rates. These can be taken to be the coefficients of the $\cos^2(\Delta
m_{\sss B} t /2)$, $\sin^2(\Delta m_{\sss B} t /2)$ and $\sin(\Delta
m_{\sss B} t)$ terms in the rates $\Gamma(\bd(t) \to (K^+\pi^-)_{\sss
  D} \ks)$ and $\Gamma(\bd(t) \to (K^-\pi^+)_{\sss D} \ks)$.

Thus, we are left with six measurements in terms of six
unknowns\footnote{In fact, it may be possible to measure $\Delta_d$ at
  a charm factory \cite{Soffer}.}: $M$, ${\overline M}$, $\beta$,
$\gamma$, $\Delta$, $\Delta_d$. Although we cannot solve the equations
analytically, once the measurements are made one will be able to
perform a fit and extract the unknown amplitudes and phases, up to
discrete ambiguities. As in the ADS modification of the GW method,
final states involving $D_{\sss CP}$ are not used.

Note that one is not constrained to use $K^+\pi^-$ as the state to
which $D^0$ and $\dbar$ decay. One can equally use another state such
as $K^+\rho^-$. In this case, only the parameter $\Delta_d$ is
changed; the remaining five parameters $M$, ${\overline M}$, $\beta$,
$\gamma$ and $\Delta$ are the same as in the $K^+\pi^-$ case. It is
therefore possible in principle to use a variety of hadronic states to
tag the $D$ mesons. By fitting to all these measurements
simultaneously, the experimental error on the CP angles can be
reduced.

In the above method, we assume that the $D$-decay amplitudes $d$ and
${\bar d}$ are known. However, if one wants to play very arcane games,
one can imagine that {\it none} of the quantities are known. If two
final hadronic states are used (say $K^+\pi^-$ and $K^+\rho^-$), then
one ends up with twelve measurements in eleven unknowns (two weak
phases, three strong phases, two $B$-decay amplitudes, four $D$-decay
amplitudes). In principle all of these unknown quantities can be
extracted from a fit to the data, up to discrete ambiguities.

The conclusion is therefore that, even in the presence of DCS
contamination, it is still possible to obtain two of the angles of
unitarity triangle, say $\beta$ and $\gamma$, from time-dependent
measurements of $\bd(t)$ into $\dbar\ks$ and $D^0\ks$. It must be
acknowledged, however, that such measurements will be difficult, and
will require $O(10^9)$ tagged $\bd$ decays. Therefore, in all
likelihood this method can only be carried out at a hadron collider.

\section{Discrete Ambiguities}

In the previous two sections we have seen that (i) one can obtain
$\sin^2(2\beta+\gamma)$ from a study of the time-dependent decays
$\bd(t)\to D^{**0} \ks$ and $\bd(t)\to {\bar D}^{**0} \ks$, and (ii)
two of the angles of the unitarity triangle can be extracted from the
rates for $\bd(t) \to (K^+\pi^-)_{\sss D} \ks$ and $\bd(t) \to
(K^-\pi^+)_{\sss D} \ks$. In this section we discuss the subject of
discrete ambiguities. Specifically, we are interested in two
questions.  First, what are the discrete ambiguities inherent in these
methods? And second, can these measurements be used to remove some of
the discrete ambiguities which remain if the CP angles are measured in
other decays?

Consider first the decays $\bd(t)\to {\bar D}^{**0} \ks$ and
$\bd(t)\to D^{**0} \ks$. From the time-dependent rates
[Eqs.~(\ref{BtoDbarKsrates1}) and (\ref{BtoDKsrates2})], one can
extract the quantities $\sin(2\Phi + \Delta)$ and $\sin(2\Phi -
\Delta)$, where $2\Phi = 2\beta + \gamma$. This means that $2\Phi$ and
$\Delta$ can be obtained with a fourfold ambiguity: if $2\Phi_0$ and
$\Delta_0$ are the true values, then the following four sets of angles
all reproduce the measured values of $\sin(2\Phi + \Delta)$ and
$\sin(2\Phi - \Delta)$:
\beq
\label{2betagammadisamb}
(2\Phi_0,\,\Delta_0) ~,~~~ (\pi+2\Phi_0,\,\pi+\Delta_0) ~,~~~
(-2\Phi_0,\,\pi-\Delta_0) ~,~~~ (\pi-2\Phi_0,\,-\Delta_0) ~.
\eeq
(As indicated in the discussion following Eq.~(\ref{twosines}), there
is also a discrete ambiguity between $\sin^2 2\Phi$ and $\cos^2
\Delta$.  This discrete ambiguity can be removed by repeating the
analysis with another $D^{**0} \ks$ final state.)

Now consider the decays $\bd(t) \to (K^+\pi^-)_{\sss D} \ks$ and
$\Gamma(\bd(t) \to (K^-\pi^+)_{\sss D} \ks$. The time-dependent rates
[Eqs.~(\ref{BtoKpiKsrates1}) and (\ref{BtoKpiKsbarrates2})] depend on
ten independent trigonometric functions of $\beta$, $\gamma$, $\Delta$
and $\Delta_d$. It is straightforward to show that these four angles
can be extracted up to a 16-fold ambiguity:
\begin{eqnarray}
\label{betagammadisamb}
\left(\beta_0,\,\gamma_0,\,\Delta_0,\,\Delta_{d0}\right) ~&,&~~~
\left(\beta_0,\,\pi+\gamma_0,\,\pi+\Delta_0,\,\Delta_{d0}\right) ~,~~~
\nonumber \\
\left(\pi+\beta_0,\,\gamma_0,\,\Delta_0,\,\Delta_{d0}\right) ~&,&~~~
\left(\pi+\beta_0,\,\pi+\gamma_0,\,\pi+\Delta_0,\,\Delta_{d0}\right) ~,~~~
\nonumber \\
\left(\pm{\pi \over 2}
+\beta_0,\,\pi+\gamma_0,\,\Delta_0,\,\pi+\Delta_{d0}\right) ~&,&~~~ 
\left(\pm{\pi \over 2}
+\beta_0,\,\gamma_0,\,\pi+\Delta_0,\,\pi+\Delta_{d0}\right) ~,~~~ 
\nonumber \\
\left(\pm{\pi \over 2}
-\beta_0,\,-\gamma_0,\,-\Delta_0,\,-\Delta_{d0}\right) ~&,&~~~ 
\left(\pm{\pi \over 2}
-\beta_0,\,\pi-\gamma_0,\,\pi-\Delta_0,\,-\Delta_{d0}\right) ~,~~~ 
\nonumber \\
\left(\pi-\beta_0,\,-\gamma_0,\,\pi-\Delta_0,\,\pi-\Delta_{d0}\right) ~&,&~~~
\left(\pi-\beta_0,\,\pi-\gamma_0,\,-\Delta_0,\,\pi-\Delta_{d0}\right) ~,~~~
\nonumber \\
\left(-\beta_0,\,-\gamma_0,\,\pi-\Delta_0,\,\pi-\Delta_{d0}\right) ~&,&~~~
\left(-\beta_0,\,\pi-\gamma_0,\,-\Delta_0,\,\pi-\Delta_{d0}\right) ~.
\end{eqnarray}

In the above, we have assumed that the bag parameter, $B_{\sss B_d}$,
is positive. Suppose, however, that this assumption is wrong, and
that, in fact, $B_{\sss B_d} < 0$. How is the above analysis affected?
As far as the weak phases are concerned, the answer is: not at all.
Changing the sign of $B_{\sss B_d}$ has the effect of changing the
signs of all the $\sin(\Delta m_{\sss B} t)$ terms in
Eqs.~(\ref{BtoKpiKsrates1})--(\ref{BtoKpiKsbarrates2}). This in turn
implies that the extracted angles will be the negatives of those
listed in the above solutions. However, note that, for every candidate
set of the weak phases $(\beta,\gamma)$, there is another (discretely
ambiguous) solution which contains the angle set $(-\beta,-\gamma)$.
In other words, as long as we have no information about the strong
phases, the extraction of weak phases is independent of the actual
sign of $B_{\sss B_d}$. (As per the discussion following
Eq.~(\ref{twosines}), this is completely analogous to what happens in
the original ADKL method.)

On the other hand, if we had some information about the strong phases,
then the actual sign of $B_{\sss B_d}$ would be important for
extracting the weak phases. For example, suppose we knew the true
value of $\Delta_d$. Then, assuming that $B_{\sss B_d} > 0$, this
method allows one to extract the CP phases up to a fourfold ambiguity
consisting of the four angle sets in the first two lines of
Eq.~(\ref{betagammadisamb}). However, if we assume instead that the
sign of $B_{\sss B_d}$ is unknown, then the four additional solutions
in the fourth line of Eq.~(\ref{betagammadisamb}) are also permitted,
leading to an eightfold ambiguity in the extraction of the weak
phases.

{}From Eq.~(\ref{betagammadisamb}), we see that, although one can
extract CP angles with these techniques, one is left with an
uncertainty due to discrete ambiguities. In fact, discrete ambiguities
plague all methods of obtaining the angles of the unitarity triangle.
This is a serious problem. There are a variety of ways of testing for
the presence of new physics in CP asymmetries: seeing if $\alpha$,
$\beta$ and $\gamma$ do indeed add up to 180 degrees, comparing
independently-measured values of the same CP angles, checking the
consistency between the measured values of these angles and the ranges
allowed by other measurements of non-CP-violating quantities, etc.\ 
\cite{Bnewphysics}. However, if there are discrete ambiguities, then
it is often the case that one of the values is consistent with the SM
(particularly when experimental error is taken into account), while
the others are not. Thus, in general, if one hopes to find new
physics, it is necessary to be able to remove the discrete ambiguities
\cite{KayLon}.

$B$-decay modes likely to be used for the extraction of $\alpha$,
$\beta$ and $\gamma$ are $\bd\to\pi^+\pi^-$, $\bd\to\Psi\ks$ and
$B^\pm\to DK^\pm$, respectively \cite{CPreview}. These decays probe
the functions $\sin 2\alpha$, $\sin 2\beta$ and $\sin^2 \gamma$ (or
equivalently $\cos 2\gamma$). Each of the three CP angles can be
obtained from these functions with a fourfold ambiguity.  However, if
one assumes that the three angles form the interior angles of the
unitarity triangle, then one is left with only a twofold discrete
ambiguity \cite{KayLon}. The form of the discrete ambiguity depends on
the signs of $\sin 2\alpha$ and $\sin 2\beta$. Denoting the true
values of the CP angles by $\alpha_0$, $\beta_0$ and $\gamma_0$, the
various twofold discrete ambiguities are summarized in Table
\ref{disambtable}.

\begin{table}
\hfil
\vbox{\offinterlineskip
\halign{&\vrule#&
 \strut\quad#\hfil\quad\cr
\noalign{\hrule}
height2pt&\omit&&\omit&&\omit&\cr
& $Sign(\sin 2\alpha)$ && $Sign(\sin 2\beta)$ && Discrete Ambiguity & \cr
height2pt&\omit&&\omit&&\omit&\cr
\noalign{\hrule}
height2pt&\omit&&\omit&&\omit&\cr
& $>0$ && $>0$ && $\left(\alpha_0, \beta_0, \gamma_0 \right) \to 
\left({\pi \over 2} - \alpha_0, {\pi \over 2} - \beta_0, \pi - \gamma_0 \right)$ & \cr
& $>0$ && $<0$ && $\left(\alpha_0, \beta_0, \gamma_0 \right) \to 
\left(-{\pi \over 2} - \alpha_0, {\pi \over 2} - \beta_0, - \gamma_0 \right)$ & \cr
& $<0$ && $>0$ && $\left(\alpha_0, \beta_0, \gamma_0 \right) \to 
\left({\pi \over 2} - \alpha_0, -{\pi \over 2} - \beta_0, - \gamma_0 \right)$ & \cr
& $<0$ && $<0$ && $\left(\alpha_0, \beta_0, \gamma_0 \right) \to 
\left(-{\pi \over 2} - \alpha_0, -{\pi \over 2} - \beta_0, -\pi - \gamma_0 \right)$ & \cr
height2pt&\omit&&\omit&&\omit&\cr
\noalign{\hrule}}}
\caption{The twofold discrete ambiguity in $(\alpha,\beta,\gamma)$ remaining 
following measurement of $\sin 2\alpha$, $\sin 2\beta$ and $\cos 2\gamma$.}
\label{disambtable}
\end{table}

Can the methods described in the previous chapters be used to remove
the final twofold discrete ambiguity? Unfortunately, the answer to
this question is no. First, as regards the method of Sec.~2, it is
obvious that $\sin^2(2\beta+\gamma)$ is the same for both angle sets
in any line of Table \ref{disambtable}. And second, for the technique
described in Sec.~3, we see that the discrete ambiguities in Table
\ref{disambtable} are among those found in the fourth line of
Eq.~(\ref{betagammadisamb}). Thus, the twofold discrete ambiguity
cannot be resolved by the $\bd\to D\ks$ studies described in Secs.~2
and 3.

Still, the methods described in the previous sections may turn out to
be useful for other reasons. First, they give independent ways of
measuring the CP angles. By comparing the values of these angles
obtained in these ways with those extracted from other decay modes, it
is conceivable that a discrepancy will be found, revealing the
presence of new physics. Second, due to penguin contributions, there
may be difficulties in measuring the angle $\alpha$ using
$\bd\to\pi^+\pi^-$ \cite{penguins}. In principle it is possible to
remove the penguin ``pollution" by either an isospin analysis
\cite{isospin} or a Dalitz-plot analysis of the decays $\bd\to\rho\pi$
\cite{Dalitz}, but these techniques are difficult as well. The methods
described above can be used to get at $\alpha$. In Sec.~2 the phase
$2\beta + \gamma = \pi + \beta - \alpha$ is probed. If $\beta$ is
known, this gives information about $\alpha$. And in Sec.~3, two
angles of the unitarity triangle can be obtained. One of these can be
taken to be $\alpha$. Furthermore, note that there is no penguin
pollution in these methods. Thus, it is possible that $\bd\to D\ks$
decays will be useful for cleanly measuring $\alpha$.

Finally, it is important to note that there is in fact a way to remove
the twofold discrete ambiguity of Table \ref{disambtable} through
measurements similar to $\bd\to D\ks$. Recently, Charles et al.\ have
proposed looking at Dalitz-plot asymmetries in the decay $\bd\to
D^\pm\pi^\mp\ks$ \cite{Charlesetal}. This final state is fed by
several intermediate resonant channels: $\bdbar \to D^+ K^{*-}$,
$\bd\to D_s^{**+} \pi^-$, and $\bd,\bdbar \to D^{**0}\ks$. The
measurement of this Dalitz-plot asymmetry enables one to extract $\sin
2(2\beta+\gamma)$. This knowledge in turn removes the discrete
ambiguity of Table \ref{disambtable}.


\section{Conclusions}

We have examined the prospects for observing CP violation in the
decays $\bd\to D\ks$, where $D$ represents $D^0$, $\dbar$, or any of
their excited states. Since $\bd$ and $\bdbar$ mesons can each decay
to both $D^0$ and $\dbar$, there are a number of different
CP-violating possibilities.

For example, $\sin^2(2\beta + \gamma)$ can be extracted from the
time-dependent rates for $\bd(t)\to {\bar D}^{**0}\ks$ and $\bd(t)\to
D^{**0}\ks$, where the $D^{**0}$ decays to $D^{(*)+}\pi^-$. This same
quantity can also be obtained using the final states $D^-\pi^+$ and
$D^+\pi^-$. However, although the branching ratio to $\dbar\ks$ is 150
times smaller that to $D^-\pi^+$, the asymmetry is 20 times bigger.
Thus, the $D^{**0}\ks$ state requires roughly a factor of three fewer
$B$'s to measure $\sin^2(2\beta + \gamma)$. Assuming $O(1)$ detection
efficiencies, we estimate that about $3\times 10^7$ tagged $B$'s are
needed to make this measurement.

In principle, the final state $\dbar\ks$ can also be used to probe
$2\beta + \gamma$. However, in practice it is very difficult to tag
the flavor of the final-state $D$-meson, so that one cannot
distinguish $\dbar\ks$ from $D^0\ks$. Nevertheless, one can obtain a
great deal of information from such decays. If one considers a
non-CP-eigenstate hadronic final state to which both $\dbar$ and $D^0$
can decay (e.g.\ $K^+\pi^-$), then one can obtain two of the angles of
the unitarity triangle from measurements of the time-dependent rates
for $\bd(t) \to (K^+\pi^-)_{\sss D} \ks$ and $\bd(t) \to
(K^-\pi^+)_{\sss D} \ks$. These measurements are admittedly difficult,
and we estimate that $O(10^9)$ tagged $\bd$ decays will be required.

Note that both of these methods are theoretically clean: there are no
penguin contributions to the decays. In addition, these two methods
give independent ways of measuring the CP angles. If one compares
these values of the angles with those extracted from other decay
modes, one may find a discrepancy. This would be a clear signal of new
physics.

Finally, suppose that $\alpha$, $\beta$ and $\gamma$ are measured via
the standard decays $\bd\to\pi^+\pi^-$, $\bd\to\Psi\ks$ and $B^\pm\to
DK^\pm$, respectively. Then, assuming that the three angles form the
interior angles of the unitarity triangle, one is still left with a
twofold discrete ambiguity. Unfortunately, the two methods described
in this paper do not resolve this discrete ambiguity. However, this
discrete ambiguity can be removed by examining Dalitz-plot asymmetries
in the $\bd\to D\ks$-like decay $\bd\to D^\pm\pi^\mp\ks$. Such
asymmetries allow one to extract, among other things, $\sin
2(2\beta+\gamma)$. Knowledge of this quantity is sufficient to remove
the remaining discrete ambiguity.

\bigskip 
\centerline{\bf Acknowledgements} 
\bigskip 
We thank Isi Dunietz and Jon Rosner for discussions. This research was
financially supported by NSERC of Canada and FCAR du Qu\'ebec.

\end{document}